%% file: main.tex
\documentclass[journal]{IEEEtran}
\usepackage{graphicx}
\usepackage{dsfont}

\ifCLASSINFOpdf
\else
\fi
\usepackage{physics}
\usepackage{pifont}
\usepackage{amssymb, amsmath,amsfonts,bm,bbm,amsthm}
\usepackage[dvipsnames]{xcolor}
\usepackage{subcaption}
\usepackage[noadjust]{cite}

\usepackage[capitalise]{cleveref}

\usepackage{algorithm}
\usepackage{algpseudocode}
\usepackage{forest}

\renewcommand{\vu}[1]{\ensuremath{\hat{#1}}}
\newcommand{\opa}{\vu{a}}
\newcommand{\opad}{\vu{a}^{\dagger}}

\newcommand{\opnn}{\vu{n}}
\newcommand{\opq}{\vu{q}}

\newcommand\ign[1]{}

\usepackage{tikz}
\usetikzlibrary{arrows.meta,calc,positioning, fit, shapes, backgrounds, shapes.geometric}
\pgfdeclarelayer{background}
\pgfsetlayers{background,main}
\newtheorem{example}{Example}

\newcommand\submittedtext{%
  \footnotesize This work has been submitted to the IEEE for possible publication. Copyright may be transferred without notice, after which this version may no longer be accessible.}%
\newcommand\submittednotice{%
  \vbox to 0pt{\vss 
  \begin{tikzpicture}[remember picture,overlay]
    \node[anchor=north, yshift=-1cm] at (current page.north) {%
      \parbox{\dimexpr\textwidth}{\centering\submittedtext}%
    };
  \end{tikzpicture}\vss}%
}

\begin{document}
%
\title{\submittednotice Secret Key Rate Analysis of Distribution Matching Algorithms for Discrete-Modulated CV-QKD}
%
%
%

\author{Micael Dias,
        Caroline Alves,
        Gabrielly Roman,
        and Søren Forchhammer, \IEEEmembership{Member,~IEEE}
\thanks{Manuscript received month XX, 2026; revised August XX, 20XX. This work was supported in part by the European Union (HORIZON-MSCA-2023 Postdoctoral Fellowship, 101153602 - COCoVaQ) and by the project \textit{Analysis and Development of Distribution Matching Algorithms for CV-QKD}, supported by QuIIN – Quantum Industrial Innovation, the EMBRAPII CIMATEC Competence Center in Quantum Technologies, with financial resources from the PPI IoT/Industry 4.0 of the MCTI, grant number 053/2023, signed with EMBRAPII.}
\thanks{Micael Dias and Søren Forchhammer are with the Department of Electrical and Photonics Engineering, Technical University of Denmark (DTU), 2800 Lyngby, Denmark. (e-mail: mandi@dtu.dk; sofo@dtu.dk).}
\thanks{Caroline Alves and Gabrielly Roman are with the QuIIN – Quantum Industrial Innovation, EMBRAPII CIMATEC Competence Center in Quantum Technologies, SENAI CIMATEC, Av. Orlando Gomes 1845, Salvador, 41650-010, BA, Brasil. (email: caroline.morais@fieb.org.br; gabrielly.roman@fbter.org.br).}}

\maketitle

\begin{abstract}
Continuous variable quantum key distribution protocols (CV-QKD) with discrete modulation have been intensively investigated to bridge the gap between ideal Gaussian modulation and modern coherent optical communication systems. To mitigate the penalty of discrete modulation, probabilistic constellation shaping (PCS) is applied to the modulation format and is typically performed by distribution matching (DM) algorithms. In this paper, we address the application of DM algorithms to perform PCS in CV-QKD protocols. We investigate the impact of approximating optimized Maxwell-Boltzman distributions with DM algorithms based on Huffman (HDM) and constant composition (CCDM) codes on the protocol's secret key rate (SKR) and tolerance to excess noise. Our results show that specifically symbol-by-symbol HDM degrades the SKR by at least 30\%, whereas CCDM matches the optimal SKR with code length of $10^3$ or more symbols. Furthermore, we also provide a statistical analysis of symbol dependence for both approaches, showing that CCDM must operate with blocks of at least $10^5$ symbols for the correlations
become negligible. Finally, we propose an algorithm to generate independent symbols following near-optimal distributions.
\end{abstract}


%
\IEEEpeerreviewmaketitle

\input{content}

\section*{Acknowledgment}

The authors would like to thank Metodi Yankov for valuable discussions on distribution matchers and system design.

\ifCLASSOPTIONcaptionsoff
  \newpage
\fi

\appendix
\input{appendix}
\bibliographystyle{IEEEtran}
\bibliography{references}
%



%






\vfill


\end{document}

%% file: content.tex
\section{Introduction}

\IEEEPARstart{Q}uantum key distribution (QKD) protocols allow the generation of secret keys with information-theoretic security against a malicious eavesdropper (Eve) \cite{usenko2025,portmann2022}. Over the years, QKD protocols have undergone continuous progress in both theory and experiment, leading to field trials with existing telecommunication infrastructure and significant steps toward commercial deployment \cite{wang2025,melgar2024,melgar2025,beppu2025}, with continuous-variable QKD (CV-QKD) protocols playing a pivotal role in this development. In these protocols, the secret key is encoded in the field quadratures of the states of light, such that it integrates naturally with conventional optical modulators and coherent detection, making compatibility seamless \cite{diamanti2015,zhang2024}. Beyond that, the application of advanced digital signal processing algorithms enables the mitigation of channel impairments, increasing the signal-to-noise ratio \cite{hajomer2024,hajomer2025a,roumestan2024}.

Although Gaussian modulated protocols benefit from more robust security proofs \cite{grosshans2002,leverrier2017,pirandola2021}, discrete modulation is less resource demanding, requiring lower random number generation rates, and improving compatibility with finite resolution digital-to-analog conversion. Discrete modulated protocols with phase-shift keying (PSK) and amplitude-phase-shift keying (APSK) modulations were proposed in \cite{zhao2009, leverrier2009b, djordjevic2019,dias2021} under the assumption of thermal-loss Gaussian channels for collective attacks, but penalties on the secret key rate (SKR) and the lack of modulation-agnostic security proofs held in depth investigations. The general SKR lower bounds and numerical methods developed in \cite{lin2019,denys2021} allowed the analysis of arbitrary modulation formats, providing the theoretical tools to apply probabilistic constellation shaping (PCS) to compensate the performance loss introduced by non-Gaussian modulation \cite{denys2021, almeida2021, wang2023, notarnicola2024}.

In classical systems, the idea of PCS is to modify the probability of constellation points to maximize mutual information, reaching the channel's capacity \cite{cho2019}. In such systems, PCS is commonly implemented by a distribution matcher (DM), an algorithmic approach to convert a bit sequence into symbols following a target distribution, which can be designed by exploiting conceptual similarities with source coding. Huffman codes were proposed for distribution matching in \cite{ungerboeck2002} and \cite{bocherer2011a} with different rules for updating the probability of the nodes in the Huffman tree, and in \cite{amjad2013a} prefix-free codes are proposed using distributions in rooted trees. In \cite{baur2014}, arithmetic source coding was used to allow dynamic symbol generation, and constant composition codes allowed the generation of typical sequences of fixed length in \cite{schulte2016}. 

Integration of distribution matching with CV-QKD systems to enable PCS has not been addressed, and existing work remains at a high level, largely analytical. In \cite{denys2021,almeida2021,wang}, QAM and APSK modulations were analyzed using binomial and Maxwell--Boltzmann (MB) distributions with a few specific optimizations under general security bounds for collective attacks, and in \cite{notarnicola2024} QAM constellations with MB shaping were optimized for the lossy channel, showing that probabilistic amplitude shaping (PAS) can be used to arbitrarily approximate channel capacity. Recently, experimental demonstrations also exhibited the effectiveness of PCS in improving performance, using MB distributions with optimized shaping parameters over QAM constellations with sizes 64, 256 and 1024 \cite{roumestan2021, roumestan2022, pan2022, pereira2022, roumestan2024, bian2025}. However, only \cite{pan2022} reported using the constant-composition distribution matcher (CCDM), while assuming perfect distribution matching. Consequently, the impact of practical finite-length distribution matching on the secret key rate remains largely unexplored. These practical effects become relevant in the finite-length regime, where DM algorithms have suboptimal performance in approximating the target distribution. In particular, the matcher's normalized divergence strongly depends on the target distribution support and the block length considered in the DM algorithm. The hardware implementation of CCDM is hindered by the inherently sequential nature of arithmetic-coding-based distribution matching and by its reliance on high-precision arithmetic operations \cite{pikus2019}. These characteristics limit parallelization, increase processing latency, and may create throughput bottlenecks in high-speed implementations \cite{cho2019}. Beyond practical issues related to matching accuracy, the statistical dependence imposed by finite block lengths is critical for applications in QKD systems. In general, DM algorithms can improve performance by considering a $k$-fold output symbol space at the cost of creating correlations for finite $k$. Specifically, CCDM imposes correlations between the output symbols because the codewords are restricted to have the same type, meaning they are not independent.

In this paper, we investigate the impact of finite-length distribution matching on probabilistically shaped discrete-modulation CV-QKD. Extending the analysis in \cite{notarnicola2024}, we evaluate the secret key rate of CV-QKD protocols with 64- and 256-QAM constellations under collective attacks with imperfect parameter estimation. Rather than assuming ideal distribution matching, as in \cite{pan2022}, probabilistic shaping is performed using the distributions actually generated by distribution matching algorithms. Since these output distributions can be derived directly from the theoretical models of the corresponding matchers, the proposed SKR analysis is entirely analytical and does not require distribution matching simulations or experimental characterization. We compare symbol-wise and pairwise Huffman and geometric Huffman matching with finite-length CCDM. Our results show that pairwise Huffman matching substantially improves the SKR over symbol-wise matching and that CCDM provides SKR close to the optimal distribution for relatively short block lengths. Finally, we quantify the statistical dependence introduced by finite-length CCDM through conditional entropy estimation and show that near-i.i.d. behavior is reached for block lengths above $10^5$.

The paper is structured as follows. In \cref{sec:cvqkd}, we review the basic structure of a CV-QKD protocol with discrete modulation and the secret key rate formula, and in \cref{sec:skr-pcs} we compute the secret key rates for optimized constellation shapes. The distribution matching algorithms are presented in \cref{sec:dm_algorithms}, and in \cref{sec:matching_accuracy} the matching accuracy is analyzed. We provide statistical analysis on the symbol dependency resulting from CCDM using conditional entropy estimation, and propose an algorithm to generate i.i.d. symbols with the same matching accuracy as CCDM. In \cref{sec:skr_results}, we evaluate the penalty of a practical distribution matcher on the secret key rate, as well as its effect on the channel excess noise tolerance. The concluding remarks are given in \cref{sec:conclusions}.

\section{System Model and Secret-Key-Rate Computation}\label{sec:cvqkd}

    We consider CV-QKD protocols with discrete modulation of the quadratures of coherent states and heterodyne detection. We use a complex-valued discrete random variable $X$ to describe the modulation scheme, called constellation, the alphabet $\mathcal{X}$ corresponds to the constellation points and has size $|\mathcal{X}|=m^2$. We restrict $X$ to be symmetric, i.e., $p(x) = p(-x)$ for all $x\in\mathcal{X}$. Equivalently, we may define two i.i.d. real-valued discrete random variables $P,Q$ such that $X = Q+iP$ and $p(x)=p(q)p(p)$. A prepare-and-measure (P\&M) CV-QKD protocol works as follows.

    \begin{enumerate}
        \item \textbf{Quantum communication} - Alice prepares coherent states $\ket{x}$ where $x$ is a realization of $X$. The corresponding ensemble is given by $\vu\rho_X = \sum_{x\in\mathcal{X}}p(x)\op{x}$. The coherent state is transmitted through a quantum channel $\mathcal{N}$ characterized by the transmittance $\tau$ and excess noise $\xi_c$. At the receiver, Bob receives the state $\mathcal{N}_{A'\rightarrow B}(\vu\rho_X)$ and performs a heterodyne measurement, whose outcomes are represented by the random variable $Y$. The measurement device has efficiency $\eta$ and electronic excess noise $\xi_d$. The procedure is repeated $L$ times and the values for $X$ and $Y$ for each round are stored in the classical registers $\mathbf{X}_L=X_1\cdots X_L$ and $\mathbf{Y}_L=Y_1\cdots Y_L$.
        
        \item \textbf{Parameter estimation} - Alice and Bob choose a random set of indexes $\mathcal{I}\subset[L]$ with $\abs{\mathcal{I}}=l$ and use $\mathbf{X}_\mathcal{I} = \qty{X_i:i\in\mathcal{I}}$ and $\mathbf{Y}_\mathcal{I} = \qty{Y_i:i\in\mathcal{I}}$ to compute the worst-case scenario $\tau_{min}$ and $\xi_{max}$ and the corresponding SKR $K$. The values used for parameter estimation are discarded, leaving the raw key sequences $\mathbf{X}_l = \mathbf{X}_L\setminus\mathbf{X}_\mathcal{I}$ and $\mathbf{Y}_l = \mathbf{Y}_L\setminus\mathbf{Y}_\mathcal{I}$. In case $K\leq0$, the protocol is aborted.

        \item \textbf{Information reconciliation} If $K>0$, Alice and Bob proceed to reconcile the raw sequences $\mathbf{X}_l\in\mathcal{X}^l$ and $\mathbf{Y}_l\in\mathbb{C}^l$. As such, Alice and Bob choose a reconciliation protocol and an error correcting code to perform error correction. After error correction, Alice and Bob sequences $S_A$ and $S_B$ are equal with high probability.

        \item \textbf{Privacy amplification} Alice and Bob apply a randomly chosen 2-universal hash function $f_H$ to $S_A$ and $S_B$ to remove both the information Eve gained during quantum communication and all the information disclosed during information reconciliation. In the end, both share the sequences $f_H(S_A)$ and $f_H(S_B)$ which are equal with high probability.        
    \end{enumerate}

    Let $\mathbb{V}(X) = \tilde{V}_m$ be the modulation variance. Then, the following quantities are parameterized by the choice of $X$,
    \begin{equation}
        \ev{n} = \tr(\opnn\vu\rho_X) = 2\bar{V}_m = \bar{m}
    \end{equation}
    \noindent is the modulation mean photon number and must be chosen to optimize the SKR, and
    \begin{equation}
        V(\opq) = \tr(\opq^2\vu\rho_X) = 4\tilde{V}_m+1 = 2\bar{m}+1
    \end{equation}
    \noindent is the variance of the quadrature operator $\opq$. 
    
    \subsection{Secret Key Rate Formula and Security Assumptions}

    We compute the SKR according to the Devetak-Winter formula for collective attacks considering finite-size parameter estimation effects \cite{devetak2005,leverrier2010b}, 
    \begin{equation}\label{eq:skr-finite}
        K=\qty(1 - \frac{l}{L})\qty(\beta I(X;Y) - \sup_{\mathcal{N}_{A\rightarrow B}}\chi_{\epsilon_{PE}}(B,E) - \Delta(L)),
    \end{equation}
    \noindent where $\beta$ is the reconciliation efficiency, $I(X;Y)$ is the classical mutual information between Alice and Bob's raw sequences, and $\chi_{\epsilon_{PE}}(B,E)$ is the upper bound on the mutual information between Eve's quantum system $E$ and Bob's random variable $Y$, considering finite block length parameter estimation with confidence interval $\epsilon_{PE}$. Information reconciliation is considered in the reverse reconciliation scheme. The supremum considers all quantum channels $\mathcal{N}_{A\rightarrow B}$ that yield the same parameters $\tau$ and $\xi_c$ obtained during parameter estimation. The fraction $l/L$ takes into account the fraction of data discarded to perform parameter estimation, and $\Delta(l)$ is the privacy amplification penalty \cite{renner2008},
    \begin{equation}
        \Delta(L) = 7\sqrt{\frac{\log2/\bar\epsilon}{L}} + \frac2L\log\frac{1}{\epsilon_{PA}},
    \end{equation}
    \noindent where $\epsilon_{PE}$ is the failure probability of parameter estimation and $\bar\epsilon$ is the smoothing parameter for the von Neumann entropy. To calculate \cref{eq:skr-finite} we use the entangled-based protocol (EB) equivalent to the P\&M previously described. The equivalent EB protocol is defined by choosing a suitable purification of $\vu\rho_X$ \cite{denys2021,ghorai2019},
    \begin{equation}
        \ket{\Psi}_{AA'} = (\mathds{1}\otimes\vu\rho_X^{1/2})\sum_{n=0}^\infty\ket{n}\ket{n}= \sum_{k=1}^N\sqrt{p_k}\ket{\psi_k}\ket{\alpha_k},
    \end{equation}
    \noindent where $\ket{\psi_k}$ is a set of orthonormal projectors on the subspace spanned by $\qty{\ket{\alpha_k}}$ such that $\vu\rho_X = \tr_A(\op{\Psi})$. Then, as Alice prepares $\ket{\Psi}_{AA'}$, mode $A$ is measured locally with $\ket{\psi_k}$, projecting the mode $A'$ into one of the coherent states in $\ket{\alpha_k}$, which is sent through the quantum channel $\mathcal{N}_{A'\rightarrow B}$. The EB and PM protocols are indistinguishable for anyone outside Alice's facilities, and the quantum state after transmission through the quantum channel is given by
    \begin{align}
        \vu\rho_{AB} &= (\mathds{1}\otimes\mathcal{N}_{A'\rightarrow B})(\op{\Psi}),\\
                     &= \sum_{j,k=1}^N\sqrt{p_jp_k}\op{\psi_j}{\psi_k}\mathcal{N}(\op{\alpha_j}{\alpha_k}).
    \end{align}

    Equivalently, Alice and Bob may consider the isometric extension $\mathcal{U}_{A'\rightarrow BE}$ of $\mathcal{N}_{A'\rightarrow B}$ such that $\vu\rho_{AB} = \tr_E[\mathcal{U}_{A'\rightarrow BE}(\op{\Psi})]$ and the same channel invariants $\tau$ and $\xi_c$. Considering $\mathcal{N}$ to be Gaussian, which is a good model for optical fibers in the linear regime, the covariance matrix for $\vu\rho_{AB}$ is
    \begin{equation}\label{eq:cov-matrix}
        \bm\Gamma = \begin{pmatrix}
            V_A \mathbb{I}_2 & Z^*(T,\xi_c)\bm\sigma_Z \\ Z^*(T,\xi_c)\bm\sigma_Z & V_B(T,\xi_c)\mathbb{I}
        \end{pmatrix},
    \end{equation}
    \noindent where $\mathbb{I}_2$ is the identity matrix of order 2, $\bm\sigma_Z=\operatorname{diag}(1,-1)$ and
    \begin{align*}
        V_A & = 2\bar{m} + 1\\
        V_B(T,\xi_c) & = 2T\bar{m} + T\xi_c + \xi_d + 1\\
        Z^*(T,\xi_c) & = 2\sqrt{T}\tr(\vu\rho_X^{1/2}\opa\vu\rho_X^{1/2}\opad) - \sqrt{2(T\xi_c + \xi_d)w},
    \end{align*}
    \noindent where $T = \tau\eta$, \ign{$\xi = T\xi_c + \xi_d$,} and
    \begin{align}
        w = \sum_{k} p_{k} \qty( \ev{\opad_{\vu\rho}\opa_{\vu\rho}}{\alpha_k} - \abs{\ev{\opad_{\vu\rho}}{\alpha_k}}^2 )
    \end{align}
    \noindent with $\opa_{\vu\rho} := \vu\rho_X^{1/2} \opa \vu\rho_X^{-1/2}$ is the first statistical moment of $\vu\rho_X$.

    In order to include the inaccuracy of parameter estimation in the system model, the worst-case values for $\tau$ and $\xi_c$ must be obtained according to the security parameter $\epsilon_{PE}$ and the sample size $l$. Here, we assume that the detection electronic noise $\xi_d$ is known from calibration and is fixed during the protocol execution. Using the normal linear model for the classical data, Alice and Bob share $y = tx + z$, where\footnote{The factor of 2 dividing the transmittance and detection noise comes from the heterodyne detection. Also, it provides 2 samples to each symbol disclosed for parameter estimation.} $t=\sqrt{T/2}$ and $z\sim\mathcal{N}(0,\sigma^2 = 1 + (T\xi_c+\xi_d)/2)$ are the effective transmittance and noise per measured quadrature, the worst-case scenario estimation values are \cite{leverrier2010b}
    \begin{align}\label{eq:tmin}
        t_{min} &= \sqrt{\frac{T}{2}} - z_{\epsilon_{PE}/2}\sqrt{\frac{1+T\xi_c/2 + \xi_d/2}{(2l)\cdot2\bar{m}}},\\\label{eq:sigmin}
        \sigma^2_{max} &= 1 + \frac{T\xi_c}{2} + z_{\epsilon_{PE}/2}\qty(1 + \frac{T\xi_c}{2} + \frac{\xi_d}{2})\sqrt\frac{1}{l},
    \end{align}
    \noindent where $z_{\epsilon_{PE}/2}$ is the $z$-coordinate so that the tail probability of a standard normal distribution is $\epsilon_{PE}/2$, i.e., $\Phi(z_{\epsilon_{PE}/2}/\sqrt{2}) = 1 - \epsilon_{PE}/2$. From \cref{eq:tmin,eq:sigmin}, one gets the corresponding minimum values for the transmittance and excess noise: $\tau_{min} = 2t_{min}^2/\eta$ and $\xi_c^{max} = 2(\sigma^2_{max} - 1 - \xi_d/2)/(\tau_{min}\eta)$. Then, we can define the covariance matrix $\bm\Gamma^{(\epsilon_{PE})}$ from which the secret key rate can be computed,
    \begin{equation}\label{eq:cov-matrix-PE}
        \bm\Gamma^{(\epsilon_{PE})} = \begin{pmatrix}
            V_A \mathbb{I}_2 & Z^*(T_{min},\xi_c^{max})\bm\sigma_Z \\ Z^*(T_{min},\xi_c^{max})\bm\sigma_Z & V_B(T_{min},\xi_c^{max})\mathbb{I}
        \end{pmatrix},
    \end{equation}
    \noindent with $T_{min}=\tau_{min}\eta$. With \cref{eq:cov-matrix-PE}, mutual information can be quickly computed using a Gaussian approximation,
    \begin{equation}
        I(X;Y) = \log(1+\frac{\tau_{min}\eta\tilde{V}_m}{2+\tau_{min}\eta\xi_{max}}).
    \end{equation}

    To compute Holevo's information, one uses the equivalence $\chi(B,E) = S(E) - S(E|B) = S(AB) - S(A|B)$, as it is assumed that Eve's system purifies Alice and Bob's joint state. The joint entropy can be obtained from the symplectic eigenvalues of $\bm\Gamma^{(\epsilon_{PE})}$, and the conditional entropy uses the symplectic eigenvalue  to Alice's state conditioned to Bob heterodyne measurement, which has 
    \begin{equation}
        \bm\Gamma_{A|B} = \qty(V_A - \frac{Z^{*2}}{V_B+1})\mathbb{I}_2.
    \end{equation}

    Holevo's information is computed according to
    \begin{align}
        \chi(B,E) &= S(AB) - S(A|B),\\
                  &= g(\nu_1)+g(\nu_2) - g(\nu_3),
    \end{align}
    \noindent where 
    \begin{equation}
        g(x) = \frac{x+1}{2}\log\frac{x+1}{2} - \frac{x-1}{2}\log\frac{x-1}{2}
    \end{equation}
    and
    \begin{align}\label{eq:symp_eigenvalue1}
        \nu_{1,2} &= [\sqrt{(V_A + V_B)^2-Z^{*2}} \pm (V_B - V_A)],\\\label{eq:symp_eigenvalue2}
        \nu_3 &= V_A - \frac{Z^{*2}}{V_B+1}.
    \end{align}

    Within the above framework, the SKR is computed under the conservative assumption that all noise observed by Alice and Bob is caused by the eavesdropper, and Holevo's information uses the total excess noise $T\xi_c+\xi_d$ to compute all eigenvalues in \cref{eq:symp_eigenvalue1,eq:symp_eigenvalue2}. Still, with the detection and channel's noise contributions being explicitly stated in the total noise expression, the parameter estimation only needs to estimate  the worst case scenarios for $\xi_c$, once detection equipments should be properly calibrated, meaning that the value of $\xi_d$ is known. This provides a hybrid approach between the known conservative and realistic assumptions \cite{usenko2016}.    

\section{Target Distribution for Probabilistic Shaping}    
\label{sec:skr-pcs}

Different probability distributions have been investigated for constellation shaping in classical and quantum communications. For the task of approaching the capacity of the Gaussian channel, special attention has been paid to families of probability distributions with good asymptotic behavior, such as the binomial and the Gauss--Hermite distributions studied in \cite{wu2010}. On the other hand, investigations of non-asymptotic scenarios for practical systems usually choose an optimized discretization of the Gaussian distribution, also known as the Maxwell--Boltzmann (MB) distribution, as it maximizes the entropy over a finite support. For QKD applications, the same constellations studied in \cite{wu2010} were analyzed in \cite{dias2021d} for unidimensional protocols. Binomial and MB distributions have also been extensively considered for square QAM modulations in CV-QKD \cite{denys2021,almeida2021,almeida2023}.

As the modulation format strongly influences the protocol SKR, the first question that has to be addressed is which modulation format provides the best SKR and, in particular, what the penalty is relative to optimal Gaussian modulation. We consider three families of probability distributions, namely, the random-walk\footnote{It has also been referred to as binomial or CLT (central limit theorem) shaping \cite{wu2010}.}, Gauss--Hermite and Maxwell--Boltzmann, defined as follows.

\begin{itemize}
    \item \textbf{Random-Walk (RW)} - define the normalized random walk random variable $R_m = \sum_{k=1}^{m-1}Z_k/\sqrt{m-1}$ with {$Z_k$} i.i.d. on $\qty{1,-1}$. Then, 
    \begin{equation}\label{eq:const-rw}
        R_m \overset{D}{=} \frac{2}{\sqrt{m-1}}\qty(B_m - \frac{m-1}{2}),
    \end{equation}        
    \noindent where $B_m\sim\operatorname{Bin}(m-1,1/2)$. Then, the RW constellation is defined by $X = Q + iP$ with $Q \overset{D}{=} P \overset{D}{=}  R_m$ and $P\perp Q$.
    
    \item \textbf{Gauss-Hermite (GH)} - For a standard Gaussian density function $p_X(x) = \frac{1}{\sqrt{2\pi}}e^{-x^2/2}$, the $m$-th Hermite polynomial given by,
    \begin{equation}
        H_m(x) = \frac{(-1)^m}{p_X(x)}\dv[m]{p_X(x)}{x}.
    \end{equation}
    The coordinates of the GH constellation in each quadrature are given by the $m$ roots $\qty{x_{i,m}}$ of $H_m$ with the corresponding weights 
    \begin{equation}\label{eq:const-gq}
        w_{i,m} = \frac{(m-1)!}{mH^2_{m-1}(x_{i,m})}.
    \end{equation}
    
    \item \textbf{Maxwell--Boltzmann (MB)} - Given the alphabet $\mathcal{X}=\pm1, \pm2, \ldots, \pm m/2$, the probability of each symbol is given by
    \begin{equation}\label{eq:mb_distribution}
        p_X(x) = \frac{e^{-\lambda x^2}}{\sum_{x'\in\mathcal{X}}e^{-\lambda x'^2}},
    \end{equation}
    \noindent with $0\leq\lambda\leq1$. 
\end{itemize}

For the RW and GH constellation formats defined above, the modulation variance $\tilde{V}_m$ can be adjusted by rescaling the amplitudes (or modifying the spacing $\delta$ between neighbor constellation points in equally spaced constellations), which is required to optimize the SKR. In the case of the MB constellation, the optimization must consider both the $\delta$ and $\lambda$ parameters. 

\begin{figure*}[!tb]
    \centering
    \begin{subfigure}{0.5\linewidth}
        \centering
        \includegraphics[width=\linewidth]{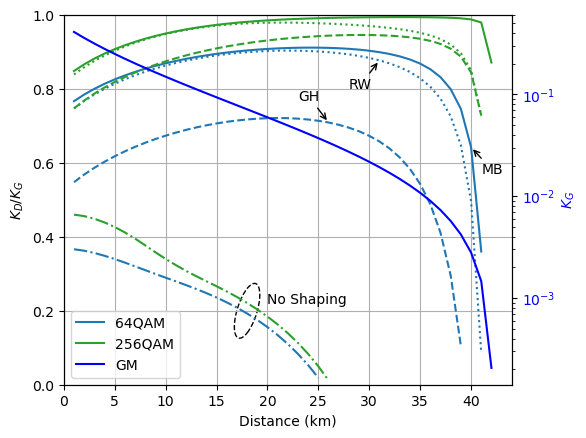}
        \caption{}
        \label{fig:SKR-PCS0.005}
    \end{subfigure}%
    \begin{subfigure}{0.5\linewidth}
        \centering
        \includegraphics[width=\linewidth]{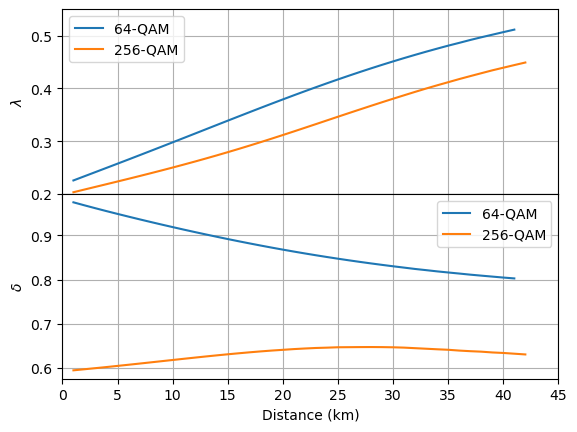}
        \caption{}
        \label{fig:MB_parameters}
    \end{subfigure}
    \caption{ (a) SKR for discrete modulation formats normalized to the Gaussian modulated protocol with optimized modulation variance at each distance point. The solid blue line corresponds to the absolute SKR value for continuous Gaussian modulation (right vertical axis). Normalized SKRs for QAM constellations of sizes 64 and 256 are represented by light blue, red and green lines. The probabilistic shaping formats MB, RW and GH are represented by solid, dotted and dashed lines, respectively. Dotted-dashed lines correspond to QAM constellations with no shaping. (b) Optimized $\lambda$ and $\delta$ parameters for the MB probabilistic shaping. The calculations considered $\beta=0.95$, detection efficiency $\eta=0.8$, fiber attenuation $0.2$dB/km and finite size parameters $L=10^{10}$, $l = L/2$, $\bar\epsilon=\epsilon_{PE}=\epsilon_{PA}=10^{-10}$, channel noise $\xi_c=0.005$ and detection noise $\xi_d=0.01$.}
    \label{fig:SKR-PCS}
\end{figure*}

In \cref{fig:SKR-PCS}, the quotient $K_{D}/K_{G}$ was plotted with key rates according to \cref{eq:skr-finite} for each constellation format, including the regular non-shaped QAM constellations, where $K_D$ denotes the SKR for discrete modulations and $K_G$ represents the SKR for a Gaussian modulation. Constellations of sizes $m^2=\qty{64, 256}$ were considered, and the modulation variance was optimized at each distance value, including a joint optimization of $\delta$ and $\lambda$ for the MB shape, which are plotted in \cref{fig:MB_parameters}. The parameters were set to $\beta=0.95$, $\eta=0.8$, fiber attenuation of $0.2$ dB/km, $L=10^{10}$, $l = L/2$, $\bar\epsilon=\epsilon_{PE}=\epsilon_{PA}=10^{-10}$, $\xi_c=0.005$ and $\xi_c=0.01$. The results indicate that probabilistic shaping enhances both the SKR and the achievable transmission distance for every constellation size. Moreover, the MB distribution with optimized parameters improves SKR up to one order of magnitude higher in relation with constellation with no shaping. The SKR penalty with respect to Gaussian modulation decreases markedly as the constellation size grows, reaching values on the order of $10^{-3}$ for 256-MB-QAM with $d>20$ km. 
 
\section{Distribution Matching Algorithms}\label{sec:dm_algorithms}

    Distribution matching refers to invertible mappings that transform uniformly distributed input bits into sequences of symbols whose distribution approximates a target distribution. In this sense, distribution matching can be regarded as the reverse operation of lossless source coding, transforming a uniform bit stream into symbol sequences that follow a target distribution rather than extracting statistical redundancy to generate nearly uniform bits. Such mappings can be realized using either variable-length prefix codes, e.g., Huffman-based shaping schemes \cite{ungerboeck2002, bocherer2011a}, or fixed-to-fixed (f2f) length matcher, e.g., CCDM \cite{schulte2016}.

    The quality of this approximation is commonly measured by the informational divergence between the output distribution of the matcher and the desired independent and identically distributed (i.i.d.) target distribution. Let $\tilde{A}^n$ denote the random output sequence produced by the matcher and let $P_A$ denote the target distribution on the output alphabet $\mathcal{A}$. The corresponding i.i.d.\ sequence is described by
    \begin{equation}\label{eq:sequence-distribution}
        P_A^n(a^n)=\prod_{i=1}^{n}P_A(a_i), \quad a^n \in \mathcal{A}^n .
    \end{equation}
    
    The divergence between the matcher output distribution $P_{\tilde{A}^n}$ and the desired distribution $P_A^n$ is defined as
    \begin{equation}
    D\!\left(P_{\tilde{A}^n}\middle\|P_A^n\right)
    =
    \sum_{a^n\in\mathcal{A}^n}
    P_{\tilde{A}^n}(a^n)
    \log_2
    \frac{P_{\tilde{A}^n}(a^n)}{P_A^n(a^n)}.
    \label{eq:kl-div}
    \end{equation}
    For f2f matchers, the full expression in \eqref{eq:kl-div} must be evaluated over sequences $a^n \in \mathcal{A}^n$, since the output symbols are not mutually independent. For a per-symbol based DM producing i.i.d. symbols according to the marginal $p$, one has that $P_{\tilde{A}^n} = p^n$ and the divergence reduces to 
    \begin{equation}\label{eq:matcher_divergence}
        D(P_{\tilde{A}^n} \| P_A^n) = nD(p \| P_A),
    \end{equation}
    which simplifies quality assessment to a per-symbol comparison between the induced marginal $p$ and the target $P_A$. The divergence on the r.h.s. of \cref{eq:matcher_divergence} is a suitable figure of merit for the matching accuracy of a given algorithm. Also, if $P_A$ is the probability distribution that maximizes mutual information for a given noisy channel, it was shown in \cite{bocherer2011} that $D(p \| P_A)$ upper bounds the capacity gap\footnote{We point out that the development in \cite{bocherer2011} consider discrete memoryless channels represented by a stochastic matrix $\qty[Q_j^i]$, but the same reasoning can be applied to continuous output memoryless channels such as the AWGN.}
    \begin{equation}\label{eq:capacity_gap_bound}
        D(p \| P_A) \geq C - \mathcal{I}(p),
    \end{equation}
    \noindent where $C$ is the channel capacity and $\mathcal{I}(p)=I(X;Y)$ is the mutual information for an input $X\sim p$ and output symbols $Y=Q(X)$, $Q$ being the channel mapping.

    \subsection{Huffman-based algorithms}\label{sec:huffman-algorithms}
    
    In Huffman-based distribution matching, a prefix-free codebook is constructed according to a target distribution defined over the constellation alphabet ($\mathcal{A}$). Each symbol $a_i$ is assigned a variable-length codeword of length $l_i$, resulting in the induced dyadic distribution $p(a_i)=2^{-l_i}$, where $l_i \in \mathbb{N}$ denotes the codeword length associated with symbol $a_i$ \cite{ungerboeck2002}. Since more probable symbols are assigned shorter codewords, the generated output sequence approximates the desired target distribution.
    
    However, the restriction to dyadic probabilities generally introduces a quantization error that prevents the resulting distribution from exactly matching the target distribution. To mitigate this effect, Huffman codes can be constructed over blocks of $K>1$ symbols. In this case, the average codeword length per source symbol is at most $1/K$ bit greater than the entropy. As a result, the rate loss tends to zero as $K \rightarrow \infty$, and the coding efficiency asymptotically approaches the entropy limit of the target distribution \cite{ungerboeck2002}. Nevertheless, classical Huffman coding does not explicitly optimize the approximation of the target distribution in terms of informational divergence. 
    
    To address this limitation, the Geometric Huffman Coding (GHC) algorithm determines the dyadic distribution that minimizes the informational divergence to a given target distribution \cite{bocherer2011a}. The difference between GHC and classical Huffman coding lies in the updating rule used during tree construction. Assuming a target distribution  $P_A = (p_1, p_2, \ldots, p_M)$ ordered in non-increasing order, $p_1 \geq p_2 \geq \cdots \geq p_M \geq 0$, whereas conventional Huffman coding combines the two least probable symbols according to
    \begin{equation}
        p' = p_{i-1} + p_i,
    \end{equation}
    GHC applies
    \begin{equation}
        p' =
        \begin{cases}
        p_{i-1}, & \text{if } p_{i-1} \ge 4p_i, \\
        2\sqrt{p_{i-1}p_i}, & \text{if } p_{i-1} < 4p_i.
        \end{cases}
    \end{equation}
     
    The particular update rule implies that classical Huffman coding minimizes the average codeword length, which is not equivalent to minimizing $D(p \| P_A)$ over the set of dyadic distributions. An illustrative example of the GHC algorithm is presented in \cite{bocherer2011a} for the target distribution

    \begin{example}\label{ex:HCxGHC}
        Consider the following distribution: $P_A = (0.328,0.32,0.22,0.11,0.022)$. Applying GHC and classical Huffman coding to this distribution yields the dyadic distributions    
        \begin{equation}
            \begin{aligned}
            p_{\mathrm{GHC}} &= (2^{-1},2^{-2},2^{-3},2^{-3},0), \\
            p_{\mathrm{HC}}  &= (2^{-2},2^{-2},2^{-2},2^{-3},2^{-3}),
            \end{aligned}
        \end{equation}
        \noindent with corresponding informational divergences    
        \begin{equation}
            \begin{aligned}
            D(p_{\mathrm{GHC}} \| P_A) &= 0.13619,\\
            D(p_{\mathrm{HC}} \| P_A) &= 0.19548.
            \end{aligned}
        \end{equation}
    \end{example}

    A fundamental characteristic of GHC is that it may assign probability zero to some symbols. This is particularly important because it may reduce the actual constellation size while increasing the mutual information, which will be explored in \cref{sec:skr_results}. Besides that, as in the case of a conventional HC, the normalized informational divergence of block-GHC-matched distributions satisfies $D(p^{(k)}\|P_A^{(k)})/k \to 0$ as $k\to~\infty$.
    
    \subsection{Arithmetic Coding and CCDM}\label{sec:ccdm-algorithm}

    Arithmetic coding (AC) is widely used for implementing distribution matchers because it enables online indexing of large codebooks without explicit storage \cite{baur2014,schulte2016}. Through a probabilistic source model, uniformly distributed input bits can be mapped to output sequences that approximate a desired target distribution \cite{baur2014,schulte2016}.

    Despite this advantage, previously proposed arithmetic-coding-based distribution matchers are variable-length schemes, which may lead to varying transmission rates, increased buffer requirements, error propagation, and synchronization issues \cite{schulte2016}. CCDM addresses these limitations through an invertible f2f mapping based on constant-composition codebooks and arithmetic coding \cite{schulte2016}. 

    The construction of CCDM is based on the concept of types. For a sequence $a^n \in \mathcal{A}^n$, the empirical distribution, or \emph{type}, is defined as

    \begin{equation}
    P_{\bar{A},a^n}(a)=\frac{n_a(a^n)}{n},
    \end{equation}

    where $n_a(a^n)$ denotes the number of occurrences of symbol $a \in \mathcal{A}$ in the sequence \cite{schulte2016}. An $n$-type is a type derived from a sequence of length $n$, and all sequences sharing the same type form a type class. A constant-composition codebook consists of codewords belonging to a common type class.

    CCDM maps $m$ uniformly distributed input bits to output sequences of fixed length $n$ belonging to a selected type class. The target composition $P_{\bar{A}}$ is chosen as the $n$-type that minimizes the informational divergence to the desired distribution $P_A$,
    \begin{equation}\label{eq:optimal_n_type}
        {P}_{\bar A} = \mathop{\mathrm{arg\,min}}\limits_{\tilde{P}} \, D(\tilde{P} \| P_A), \quad \tilde{P} \text{ is an } n\text{-type}.
    \end{equation}
    \noindent which can be obtained with the optimal quantization algorithm proposed in \cite{bocherer2016}. Let $\mathcal{T}_{{P}_{\bar A}}^{n}$ denote the type class associated with ${P}_{\bar A}$. The CCDM encoder $f_{\mathrm{ccdm}}:\{0,1\}^{m}\rightarrow\mathcal{T}_{{P}_{\bar A}}^{n}$ uses    
    \begin{equation}
        m= \left\lfloor\log_2\left|\mathcal{T}_{{P}_{\bar A}}^{n}\right|\right\rfloor
    \end{equation}
    to ensure that the mapping is invertible \cite{schulte2016}. The normalized divergence between the CCDM output distribution and the target memoryless source is given by \cite{schulte2016}
    \begin{equation}
        \frac{1}{n}D\!\left(P_{\tilde{A}^n}\,\middle\|\,P_A^n\right)=
        H(\bar{A})-R+D({P}_{\bar A}\|P_A),
        \label{eq:ccdm-div}
    \end{equation}
    \noindent where $R=m/n$ is the matching rate. The term $D({P}_{\bar A}\|P_A)$ accounts for the approximation of the target distribution by the selected $n$-type, whereas $H(\bar A)-R$ corresponds to the finite-length rate loss. As the block length increases, $\bar P_A$ converges to $P_A$, $H(\bar A)$ approaches $H(A)$, and the matching rate approaches the entropy of the target distribution. Consequently, the normalized divergence vanishes asymptotically \cite{schulte2016}.   

    \subsection{Implications for Reverse Reconciliation and MD Reconciliation}\label{sec:dm_ir}
    
    In classical communication systems, the distribution matcher operates jointly with forward error correction (FEC) to maximize the information rate, and this integration has been extensively studied \cite{bocherer2015,cho2018,gultekin2020}. However, in CV-QKD, information reconciliation differs from a classical decoding because the receiver does not need to recover the transmitted information: Alice and Bob just need to share a pair of identical binary sequences. Moreover, reverse reconciliation (RR) is typically applied to overcome the 3dB loss limit, and the multidimensional reconciliation is used to achieve reconciliation efficiency above $0.95$ in the low SNR regime. Consequently, the integration of distribution matchers into the system architecture raises questions that are absent in its applications to classical communication systems.
    
    On a theoretical level, the first issue comes from an overlooked mismatch between MD reconciliation and discrete modulation. The idea behind MD reconciliation is to exploit the fact that normalized $d$-dimensional Gaussian vectors are uniformly distributed over the unit sphere $\mathcal{S}^{d-1}$ for $d=\qty{1,2,4,8}$, which holds naturally for Gaussian-modulated protocols with homodyne or heterodyne measurements but does not generally hold for discretely modulated systems. Although the channel output observed by Bob may approach Gaussian statistics under sufficiently noisy conditions, the non-Gaussian nature of Alice's raw symbols prevents a direct extension of the original arguments. Despite this theoretical mismatch, MD reconciliation has been applied to CV-QKD protocols with discrete modulation because it performs well in practice, providing reconciliation efficiencies above $0.95$ \cite{lu2025,gumus2025,almeida2023}.
    
    A more significant consequence of employing the MD protocol with reverse reconciliation is that no dematching operation is required at Bob's side. In a classical communication system with distribution matching, the received symbols must be mapped back to the corresponding binary sequence, which is performed jointly with the ECC decoder. In a CV-QKD system with MD reconciliation, the secret key bits are generated locally on the reference side (Bob, in the case of RR), and the channel output is used to calculate the necessary rotations on $\mathcal{S}^{d-1}$, so that the secret key values and Alice's vector are correlated, allowing a channel coding treatment. Moreover, variable length DM algorithms such as Huffman and arithmetic codes lead to issues such as varying transmission rate, error propagation and synchronization problems in classical communication system. For applications in CV-QKD, these issues can be handled using some sort of many-to-one strategy, such as in \cite{yankov2016}, but they can be completely avoided if no dematching operation is needed at Bob's side. Even though length-related issues can be avoided using an f2f code, not performing a dematching operation provides a seamless integration with the MD setup while synchronization or catastrophic error-propagation issues are also avoided.

\section{Distribution Matching Accuracy}\label{sec:matching_accuracy}

    As presented in \cref{sec:skr-pcs}, among the probabilistic shaping formats that were considered, the optimized MB distribution provides the highest SKR for QAM constellations, making it the optimal distribution to be targeted by the DM. Moreover, integration of distribution matching, reverse reconciliation, and multidimensional reconciliation exempts the use of a dematcher on Bob's side, as discussed in \cref{sec:dm_ir}, and the task of the DM is to accurately synthesize the optimal MB output distribution. In this Section, we analyze the approximated distribution effect from a classical information perspective.
    
    Since the constellations in \cref{fig:SKR-PCS} consider probability distributions with i.i.d. marginals, it is sufficient to evaluate the matching accuracy per dimension. Let $\mathcal{X}=\qty{\delta(k-(m-1)/2): k=0,\cdots, m-1}$ be, with $m=\qty{8,16}$, the set of amplitudes per dimension in a QAM constellation with 64 or 256 points, respectively, and $\delta>0$. For each pair $(\lambda,\delta)$ in \cref{fig:MB_parameters}, we calculate the divergence $D(p_{\tilde{X}}(x;\lambda)\|p_X(x;\lambda))$, where $X\sim p_X(x;\lambda)$ is given by \cref{eq:mb_distribution} and $p_{\tilde{X}}(x;\lambda)$ is the corresponding distribution output of the DM. We consider symbol-by-symbol Huffman matchers (HDM and GHDM) as well as a higher dimensional pairwise encoding, denoted by 2HDM and 2GHDM. The super-symbols follow a product distribution $(X_1,X_2)\sim p_{X_1}(x_1;\lambda)\cdot p_{X_2}(x_2;\lambda)$ targeted by the DM. For the CCDM, we considered the output with lengths $n=\qty{10^2, 10^3}$.
    
    \begin{figure}[!tb]
        \centering
        \begin{subfigure}{\linewidth}
            \includegraphics[width=\linewidth]{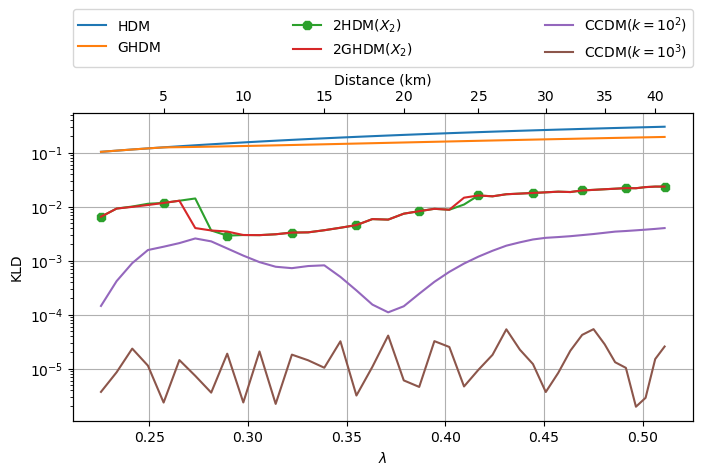}
            \caption{$m=8$}
            \label{fig:KLD-DG-match8}
        \end{subfigure}
        \begin{subfigure}{\linewidth}
            \includegraphics[width=\linewidth]{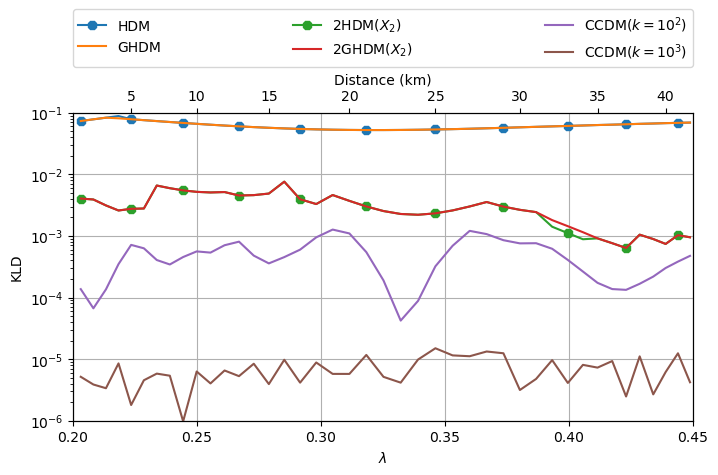}
            \caption{$m=16$}
            \label{fig:KLD-DG-match16}
        \end{subfigure}
        \caption{Kullback–Leibler divergence between matched distributions and the optimized MB distribution supports of size $m=\qty{8,16}$ for the Huffman matchers (symbol and pairwise) and the CCDM with different block of lengths.}
        \label{fig:matching_accuracy}
    \end{figure}

    \Cref{fig:matching_accuracy} shows the divergences for MB distributions with 8 and 16 points as a function of $\lambda$. To provide a direct connection with the key rates of \cref{fig:SKR-PCS0.005}, the upper horizontal axis contains the distance values corresponding to each $\lambda$. For the 8-point distribution, both HDM and GHDM exhibit divergences over $10^{-1}$ per dimension, with GHDM consistently providing a slight improvement over HDM, and increasing the Huffman output dimension to $n=2$ reduces the divergence by approximately one order of magnitude. Moreover, the performance of HDM and GHDM becomes nearly indistinguishable in this regime. This occurs because $HC(p_X(x;\lambda)) = GHC(p_X(x;\lambda))$ for most pairs $(\lambda,\delta)$. CCDM achieves significantly lower divergences than the Huffman approaches, below $10^{-2}$ for $n=10^2$ and on the order of $10^{-5}$ bits per dimension for $n=10^3$. Similar behavior is observed for the 16-point distribution, except that HDM and GHDM behave nearly equal for the entire range of $\lambda$ for symbol by symbol and pairwise matching. 

    As discussed previously, the relative entropy provides a measure of matching accuracy, which can be related to the classical capacity gap (\cref{eq:capacity_gap_bound}), and the divergence values presented in \cref{fig:matching_accuracy} show the divergence mismatch of up to $10^{-5}$ bits for CCDM. Although the distributions in \cref{fig:SKR-PCS} were not optimized to maximize mutual information (but for SKR, which combines classical and quantum information), it is reasonable to assume that DM algorithms that provide lower divergences also result in lower SKR penalties. For example, in \cite{dias2021d} it was shown that the SKR gap of capacity achieving distributions follows the same trends as the classical gap investigated in \cite{wu2010} for the Gaussian channel.

    \subsection{Pairwise Huffman Matching Asymmetry}

\begin{figure}
        \includegraphics[width=\linewidth]{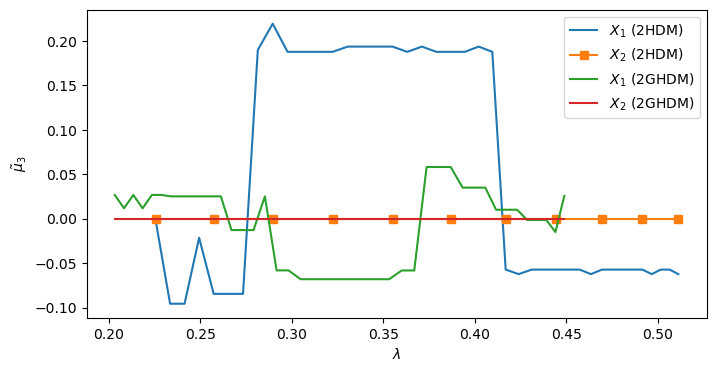}
        \caption{Skewness of the marginal distributions obtained from pairwise Huffman and geometric Huffman matching from \cref{fig:KLD-DG-match8}.}
        \label{fig:skewness}
    \end{figure}

     Pairwise Huffman matching improves the matching accuracy by at least one order of magnitude relative to a symbol-wise approach, but it comes at the cost of the symbols $(X_1,X_2)$ output by the matcher being inherently correlated. Since the QKD protocol assumes that the source symbols are i.i.d., a trivial solution is to discard the symbol with the largest divergence from the optimal MB distribution. However, one can verify that the marginals $p_{\tilde{X}_1}(x;\lambda)$ and $p_{\tilde{X}_2}(x;\lambda)$ are not identically distributed and are not necessarily symmetric. To quantify the asymmetry of the marginals, we evaluated the skewness $\tilde{\mu}_3 = \mathbb{E}[(X-\mathbb{E}X)/\sigma_X]^3$ for both marginals and plotted it in \cref{fig:skewness}. The results show that $\tilde{X}_1$ is asymmetric for all $\lambda$, while $\tilde{\mu}_3(\tilde{X}_2)=0$. Although zero skewness is not a sufficient condition for symmetry, direct inspection confirmed that $p_{\tilde{X}_2}(x;\lambda) = p_{\tilde{X}_2}(-x;\lambda)$ for every $x\in\mathcal{X}$.

    Therefore, the correlation and asymmetry inherent in pairwise Huffman DM can be avoided by discarding the first symbol and performing SKR calculations using only $p_{\tilde{X}_2}(x;\lambda)$. Analogously to the Huffman-based algorithms, CCDM also introduces correlations due to the fixed-type, constant-composition constraint. A dedicated correlation analysis is presented in \cref{sec:ccdm_correlation}.

    \subsection{Symbol Dependence in CCDM}\label{sec:ccdm_correlation}

    As discussed in \cref{sec:ccdm-algorithm}, the underlying idea of CCDM is to restrict the output sequences to have the same type, forming a constant composition code. This restriction introduces unavoidable correlations between the symbols within the codeword, breaking the i.i.d. hypothesis of QKD. For example, take the binary sequences of type $n=5$ with $p(0) = 3/5$ and $p(1) = 2/5$. The second bit is conditionally dependent on the first, $p(1|0) = 1/2$ and $p(1|1) = 1/4$, therefore, is not independent, and the conditional entropy $H(B_2|B_1)$ is reduced by $0.046$ bit compared to $H(B_2)$.

    Finite length operation is more difficult to analyze if one aims to understand symbol dependence in CCDM. Unlike Huffman matching, whose output joint and marginal distributions are known in advance for any block length, having a closed-form model for the CCDM conditional distribution is intractable for large $n$. Then, we analyze sequential correlations through statistical estimation of the conditional entropy $H(X_i|X_{i-1}\cdots)$ for several block lengths. More specifically, we want to determine how the observation of a short sequence of symbols output by the CCDM changes the uncertainty on the next one. Although this strategy imposes a simplification by considering only a short memory and discarding possible correlations with the remaining sequence, it can be useful to grasp the convergence of conditional entropy towards $H(P)$, implying independency.

    The random experiment is defined as follows. Fix a block length $n$ and a target distribution $P_A$. The corresponding $n$-type distribution $P_{\bar{A}}$ is also fixed. A random sequence of $k$ i.i.d. bits is drawn and input to the CCDM, producing the sequence $X_1,\cdots,X_n$. We estimate the sample's joint distribution $\hat{p}(i,j) = n(i,j)/N$, where $n(i,j) = \sum_{l=2}^k\mathbb{I}(X_l=i,X_{l-1}=j)$, and compute its conditional entropy $H_{c_i}(X_i|X_{i-1})$. The experiment is repeated $N$ times, and the estimated conditional entropy is the averaged sample conditional distribution
    \begin{equation}\label{eq:entropy_estimator}
        \hat{H}(X_i|X_{i-1}) = \frac1N\sum_{i=1}^NH_{c_i}(X_i|X_{i-1}).
    \end{equation}

\begin{figure}
        \centering
        \includegraphics[width=1\linewidth]{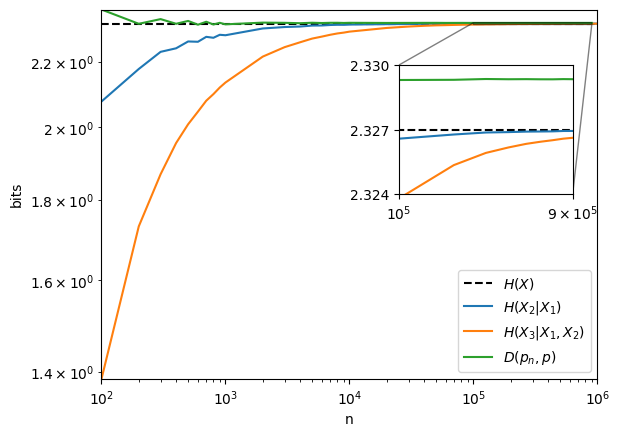}
        \caption{Conditional entropy $\hat{H}(X_i|X_{i-1})$ and $\hat{H}(X_i|X_{i-1},X_{i-2})$ for the sequences output by the CCDM with different block lengths.}
        \label{fig:entropies}
    \end{figure}

    \cref{eq:entropy_estimator} can be extended to $\hat{H}(X_i|X_{i-1}, \cdots, X_{i-k})$. \cref{fig:entropies} shows the estimates for $H(X_i|X_{i-1})$ and $H(X_i|X_{i-1}, X_{i-2})$ for $P_A = (0.00537, 0.04019, 0.15374, 0.30069, 0.30069, 0.15374,\allowbreak 0.04019, 0.00537)$, block lengths ranging from $10^2$ to $10^6$, and $N=10^3$. In Figure \ref{fig:entropies}, both conditional entropies start far apart for $n<10^3$ and converge to $H(X)\approx 2.327$ bits. As expected, increasing the conditioning sequence reduces the uncertainty over $X_i$, making $\hat{H}(X_i|X_{i-1})\leq\hat{H}(X_i|X_{i-1},X_{i-2})\leq H(X)$, getting close to equality for $n>10^5$, implying practical independence.
    
    Entropy estimation is beyond the scope of this paper. The estimator in \cref{eq:entropy_estimator} is the plug-in (maximum-likelihood) estimator, whose finite-sample properties are well known \cite{paninski2003}. In particular, it exhibits a negative bias that increases as the state space grows relative to the sample size. As a result, the conditional entropy estimates in \cref{fig:entropies} are expected to lie below their true values, with the effect being stronger for joint distributions than for the marginal entropy $H(X)$. The observed gap between the conditional entropies and $H(X)$ at short block lengths therefore reflects both genuine dependence and estimation bias. Since averaging over $10^3$ realizations reduces variance but not bias, the convergence of the conditional entropies to $H(X)$ around $n\approx10^5$ supports the conclusion that correlations are negligible for blocks of this length. For shorter blocks, however, estimator bias should be taken into account when interpreting the apparent dependence and assessing practical security.

    \subsection{Many-to-one Huffman Distribution Synthesis}

    Using higher dimensions is the natural strategy to improve matching accuracy, as observed in the divergence plots of \cref{fig:matching_accuracy}. Pairwise Huffman matching reduces divergence by one order of magnitude, and CCDM provides even better improvements with relatively short block lengths because its output distribution is chosen to be the closest $n$-type approximation to the target distribution (in the sense of informational divergence). Such improvements come with a price, as block-wise matching imposes unavoidable correlations between symbols. In the case of pairwise Huffman matching, discarding one symbol of each pair is sufficient to ensure the i.i.d. structure required by the protocol, but CCDM spreads correlations along the entire block, making them more difficult to track. To overcome this issue, we propose a many-to-one Huffman distribution-synthesis solution, whose idea is to adapt the optimal $n$-type distributions such that they can be synthesized using a symbol-wise Huffman tree. With this approach, one can obtain the same distributions as in CCDM while avoiding correlations, since the Huffman tree is designed to output single symbols. The description of the algorithms is given in the Appendix.

\section{Secret Key Rates and Excess Noise Tolerance}\label{sec:skr_results}

\begin{figure}[!tb]
        \centering
        \begin{subfigure}{\linewidth}
            \centering
            \includegraphics[width=\linewidth]{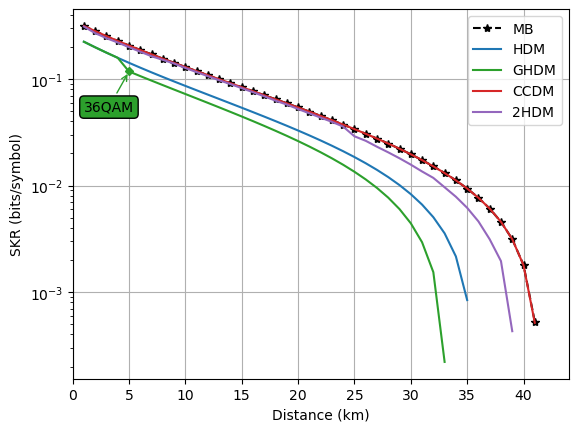}
            \caption{64-QAM}
            \label{fig:skr-DG-DM-64QAM}
        \end{subfigure}
        \begin{subfigure}{\linewidth}
            \centering
            \includegraphics[width=\linewidth]{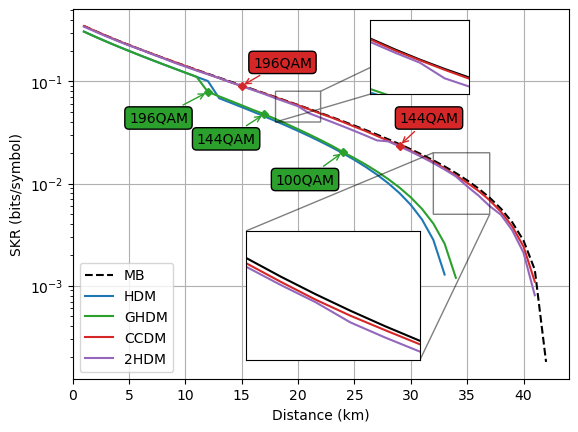}
            \caption{256-QAM}
            \label{fig:skr-DG-DM-256QAM}
        \end{subfigure}
        \caption{Secret key rate for probabilistically shaped (a) 64-QAM and (b) 256-QAM constellations considering the optimal MB distributions and the matched distribution from the Huffman and CCDM approaches.}
        \label{fig:skr-DG-DM}
    \end{figure}

    The SKR is a function of the modulation format given a fixed set of physical and security parameters. As in \cref{sec:matching_accuracy}, we start with the optimal shaping pair $(\lambda, \delta)$ from \cref{fig:MB_parameters} and calculate $p_{\tilde{X}}(x)$ to compute the SKR of each DM algorithm. A second optimization of $\delta$ is performed to ensure a maximal SKR for the DM distribution. The results are shown in \cref{fig:skr-DG-DM-64QAM,fig:skr-DG-DM-256QAM} for shaped 64-QAM and 256-QAM, respectively. The plots for the block-Huffman DM consider only the distribution of the symbol $X_2$, as discussed in \cref{sec:matching_accuracy}, and the CCDM output $n$-type distribution has fixed $n=10^3$.

    For 64-QAM, HDM and GHDM have poor performance, exhibiting lower SKR compared to the optimal MB shape. Specifically, GHDM assigns probability zero to 28 symbols, reducing the effective constellation size to 36 points for $d\geq5$, resulting in an SKR lower than HDM. The SKR for 2HDM matches the optimal MB PCS for $d<25$ km and drops beyond that. CCDM provides the best SKR, matching the optimal MB shape for the entire distance range.

    A similar behavior is exhibited for the 256-QAM constellations. GHDM also reduces the effective constellation size by assigning probability zero to some constellation symbols, yielding sizes 196, 144 and 100 as the distance increases. Unlike the 64-QAM case, the SKR for the constellations using GHDM is not critically penalized and is almost the same as for 256-QAM with HDM shaping. The 2HDM also exhibits SKR close to the optimal MB shaping for almost all distances, with a visible gap for a few distance values. CCDM again provides the best matching accuracy, exhibiting the highest SKR. However, for 256-QAM, the support of the optimal $n$-type distribution is smaller than the support of the optimal MB distribution for several distances. The effective constellation sizes are 196 for $15\leq d\leq 28$ and 144 for $d>28$.

    The maximal excess noise that a QKD protocol tolerates is an important figure of merit, as it is directly related to operational boundaries and applications in real-world scenarios. Since, in general, the detection noise does not depend on the channel conditions, we evaluate the maximum channel excess noise $\xi^*_{c}$, defined as the largest value of $\xi_c$ for which the optimized SKR is no lower than the threshold $K_{th}=10^{-4}$
    \begin{equation}\label{eq:max_excess_noise}
        \xi^*_c = \sup\qty{\xi_c: \max_{\lambda,\delta}K(\xi_c, \lambda,\delta)\geq K_{th}},
    \end{equation}
    \noindent where $K(\xi_c, \lambda,\delta)$ is the SKR function for the MB shaped constellation. Numerically, it is equivalent to find the root of
    \footnote{
    Since the SKR is a monotonic decreasing function of $\xi$, for every $\xi_1\geq\xi_2$, $K(\xi_1)\geq K(\xi_2)$, so that the feasible set for \cref{eq:max_excess_noise} is $[0,\xi_c^*]$ and the boundary satisfies $\max_{\lambda,\delta}K(\xi^*_c, \lambda,\delta)=K_{th}$.
    }
    \begin{equation}\label{eq:excess_noise_root}
        f(\xi_c) = \max_{\lambda,\delta}K(\xi_c, \lambda,\delta) - K_{th},
    \end{equation}
    \noindent which can be solved with Brent's root finding algorithm \cite{brent2013}. The same formulation can be used for the SKR considering DM with the modification that the maximum optimizes only over the $\delta$ parameter.

\begin{figure}[!tb]
        \centering
        \begin{subfigure}{\linewidth}
            \centering
            \includegraphics[width=\linewidth]{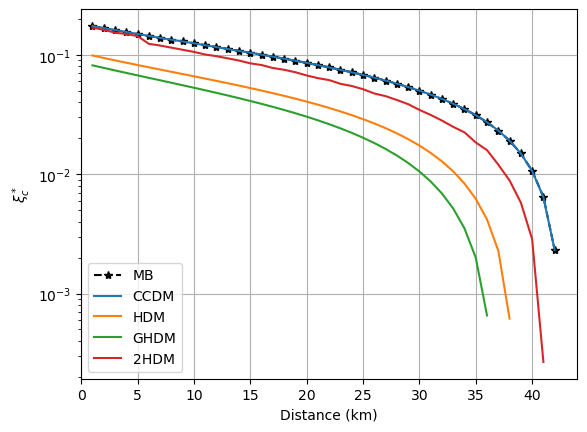}
            \caption{64-QAM}
            \label{fig:max_noise_64}
        \end{subfigure}
        \begin{subfigure}{\linewidth}
            \centering
            \includegraphics[width=\linewidth]{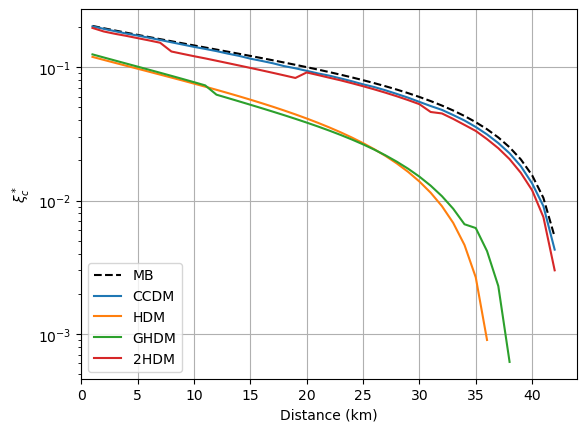}
            \caption{256-QAM}
            \label{fig:max_noise_256}
        \end{subfigure}
        \caption{Maximal Excess Noise for a SKR threshold of $K_{th}=10^{-4}$ of probabilistically shaped (a) 64-QAM and (b) 256-QAM constellations considering the optimal MB distributions and the matched distribution from the Huffman and CCDM approaches.}
        \label{fig:max_noise}
    \end{figure}

    \cref{fig:max_noise} shows the values of $\xi^*_{c}$ for 64-QAM and 256-QAM considering different DMs. The results for the 64-QAM case show the same trend as \cref{fig:skr-DG-DM} for the maximal SKR, except that there is almost no overlap between the values of $\xi^*_{c}$ for different DM algorithms. Specifically, CCDM has the same noise tolerance as the optimal MB, followed by 2HDM, HDM and GHDM. For 256-QAM, $\xi^*_{c}$ behaves similarly, with CCDM providing the highest values of $\xi^*_{c}$, within $10^{-3}\sim10^{-2}$ SNU of the optimal MB shaping.

\section{Conclusion}\label{sec:conclusions}

In this paper, we investigated how finite-length distribution matching impacts the performance of discretely modulated CV-QKD with probabilistic constellation shaping. Focusing on the Maxwell--Boltzmann (MB) distribution that maximizes the secret key rate (SKR), we compared distribution matching algorithms based on Huffman coding and constant-composition distribution matching (CCDM) in terms of matching accuracy, achievable SKR, and excess-noise tolerance.

Our results show that CCDM provides the most accurate approximation of the target MB distribution and consequently achieves the highest SKRs, essentially matching the MB benchmark across the considered distances for both 64-QAM and 256-QAM. Pairwise Huffman matching (2HDM/2GHDM) can approach the MB performance over a significant range, but introduces symbol correlations and marginal asymmetry, which can be mitigated by using only the symmetric marginal (\cref{sec:matching_accuracy}). Symbol-by-symbol Huffman matching exhibits the largest divergence from the target distribution and, in the case of GHDM, may assign zero probability to several constellation points, effectively shrinking the constellation and reducing the SKR.

The same trends are reflected in the maximum tolerable channel excess noise, where CCDM yields the best tolerance, followed by pairwise Huffman matching and then the symbol-by-symbol variants (\cref{fig:max_noise}). Finally, although CCDM necessarily breaks the i.i.d. assumption at finite block length, we conducted a statistical analysis of conditional entropies, indicating that these correlations become negligible in practice for sufficiently long blocks ($n\geq 10^5$), supporting the use of CCDM in shaped CV-QKD implementations.

%% file: appendix.tex
\label{sec:mto_huffman}
 
    In this section, an alternative method based on the construction of an extended dyadic tree is presented, which allows for the independent generation of symbols according to a $q(x)$ distribution of type $M$, chosen as an approximation of a target distribution $p(x)$, with $M = 2^k$. The main idea consists of extending the original alphabet of the distribution of type M and defining a function that maps the symbols of the distribution of type M to the extended distribution. The main characteristic of the extended distribution is that it can be represented by the leaf nodes of a Huffman tree, and consequently by a prefix code.
    
    The construction starts from the representation $q(x) = a_x / M$, with $a_x \in \mathbb{Z}_{\geq 0}$ and $\sum_x a_x = M$. Each coefficient $a_x$ is then decomposed into its binary expansion,
    \begin{equation}
    a_x = \sum_i 2^{b_{x,i}},
    \end{equation}
    which induces a decomposition of the probability $q(x)$ in dyadic terms of the form $2^{-l_{x,i}}$, with $l_{x,i} = k - b_{x,i}$.
    
    This decomposition defines an extended alphabet $\mathcal{X}'$, composed of subsymbols $(x,i)$ with probabilities $2^{-l_{x,i}}$. Next, a Huffman tree is constructed over $\mathcal{X}'$, resulting in a complete prefix code whose lengths coincide with the values $l_{x,i}$. As a consequence, multiple leaves of the tree can be associated with the same original symbol $x$.
    
    \begin{example}\label{ex:huffman-estendido}
        Consider the distribution $p=(3/8,3/8,1/8,1/8)$ over the alphabet $(a,b,c,d)$. \Cref{fig:exemplo-huffman} shows the Huffman code for $p$. The extended distribution $p^+=(2/8,1/8,2/8,1/8,1/8,1/8)$ can be obtained from the Huffman code from \Cref{fig:exemplo-huffman-estendido}.
    \end{example}
    
    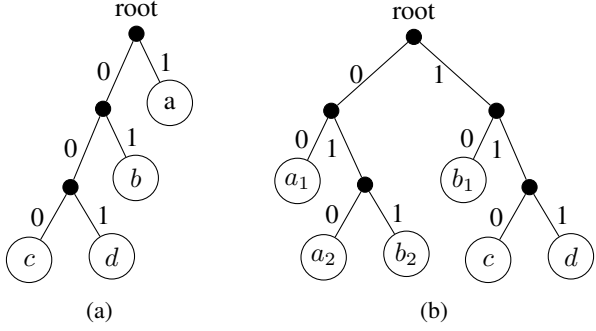
\begin{figure}[t]
        \input{Figures/huffman_trees}
    \caption{(a) Huffman tree and (b) extended Huffman tree for the the example distribution $p=(3/8,3/8,1/8,1/8)$}
    \end{figure}

%% file: Figures/huffman_trees.tex
\begin{subfigure}{0.4\linewidth}
    \centering
    \begin{forest}
        for tree={
            draw,
            inner sep=1pt,
            s sep=5mm,
            l sep=5mm,
            circle,
            if n children=0{minimum size=6mm}{fill=black,minimum size=2mm,}               
        }
        [, label=above:{root}
            [ , edge label={node[midway,left]{0}}
                [, edge label={node[midway,left]{0}}
                [$c$ , edge label={node[midway,left]{0}}]
                [$d$ , edge label={node[midway,right]{1}}]
                ]
                [$b$ , edge label={node[midway,right]{1}}]
            ]
            [a , edge label={node[midway,right]{1}}                
            ]
        ]
    \end{forest}
    \caption{}
    \label{fig:exemplo-huffman}
\end{subfigure}%
\begin{subfigure}{0.6\linewidth}
    \centering
    \begin{forest}
        for tree={
            draw,
            inner sep=1pt,
            s sep=5mm,
            l sep=5mm,
            circle,
            if n children=0{
                minimum size=6mm,
            }{
                fill=black,
                minimum size=2mm,
            }               
        }
    [, label=above:{root}
        [ , edge label={node[midway,left]{0}}
            [$a_1$ , edge label={node[midway,left]{0}}]
            [ , edge label={node[midway,left]{1}}
                [$a_2$ , edge label={node[midway,left]{0}}]
                [$b_2$ , edge label={node[midway,right]{1}}]
            ]
        ]
        [ , edge label={node[midway,left]{1}}
            [$b_1$ , edge label={node[midway,left]{0}}]
            [ , edge label={node[midway,left]{1}}
                [$c$ , edge label={node[midway,left]{0}}]
                [$d$ , edge label={node[midway,right]{1}}]
            ]
        ]
    ]
    \end{forest}
    \caption{}
    \label{fig:exemplo-huffman-estendido}
\end{subfigure}